\documentclass[a4paper]{article}

\usepackage{graphicx}
\usepackage{pdfpages}
\usepackage{mathtools}
\usepackage{multirow}

\providecommand{\keywords}[1]
{
  \small	
  \textbf{\textit{Keywords---}} #1
}

\usepackage[sorting=none]{biblatex}
\begin{document}

\title{Short note: Transformation between different solution methods for general axisymmetric tangential contact problems in Hertz-Mindlin approximation}

\author{E. Willert \\
Technische Universit\"at Berlin\\
Stra\ss{}e des 17. Juni 135, 10623 Berlin, Germany \\
e.willert@tu-berlin.de}
\maketitle

\begin{abstract}
The transition between two conceptionally different solution procedures for general axisymmetric tangential contact problems with arbitrary laoding histories under Hertz-Mindlin assumptions is demonstrated, namely J\"ager's superposition solution and the method of dimensionality reduction. Both finite and infinite superpositions of Cattaneo-Mindlin basis functions are considered. It is shown how the weights in the superposition solution can be easily obtained from the displacements in the MDR model.
\end{abstract}


\section{Introduction}\label{sec1} 

There are several different semi-analytical solution procedures to axisymmetric tangential contact problems with arbitrary loading histories under Hertz-Mindlin assumptions (neglect elastic coupling, uni-directional tangential tractions, neglect lateral displacements), for example the original algorithm by Mindlin \& Deresiewicz \cite{MindlinDere1953}, which was later simplified by Jäger \cite{Jaeger1993} – who demonstrated that the solution for arbitrary loading can always be written in terms of a finite or infinite superposition of the Cattaneo-Mindlin basis solution for the elementary loading procedure of indentation followed by tangential displacement without reversing the direction of tangential motion - the method of memory diagrams (MMD) by Aleshin \textit{et al.} \cite{AleshinAbeele2012} or the method of dimensionality reduction (MDR) by Popov \textit{et al.} \cite{PopovHess2014}. Whereas all solution methods are (within the Hertz-Mindlin approximation), of course, equivalent to each other, they have different "perks", i.e. for specific problems one may be more appropriate or easier to implement than the others, and under some circumstances it may be desirable to be able to switch between the different solution formulations. In the following the transformation between the MDR solution and Jäger’s superposition is demonstrated.

\section{Transformation procedure}\label{sec2}

\subsection{The Cattaneo-Mindlin basic solution}\label{sec2_1}

Consider the following elementary loading of the contact between a rigid indenter with the smooth axisymmetric profile $f(r)$, $r$ being the polar radius in the contact plane, and an elastic half-space with the shear modulus $G$ and Poisson number $\nu$ (the generalization to contacts of two elastic bodies is trivial under the assumptions stated above \cite{Popovetal2019}): first the indenter is pressed into the half-space until a contact radius $a$ is reached and subsequently displaced in the tangential $x$-direction by $u_0 > 0$. The coefficient of friction is $\mu $. According to the principle by Ciavarella (\cite{Ciavarella1998b}, \cite{Ciavarella1998}) and Jäger \cite{Jaeger1998} the tangential tractions in the contact area as a function of the radial coordinate, within the framework of the Hertz-Mindlin approximation, can be written as
\begin{align}
\sigma_{xz}(r) \coloneqq \tau (r) = \tau_B(r;a,c) \coloneqq \mu \begin{cases}
p(r;a) - p(r;c) 	&	\text{if} \quad r \le c \\
p(r;a)					&	\text{if} \quad c < r \le a \\
0,					& \text{else}
\end{cases}
\end{align}
with the stick radius $c$ (which can be easily determined from $u_0$ \cite{Popovetal2019}) and the pressure distribution $-\sigma_{zz}(r;\tilde{a}) \coloneqq p(r;\tilde{a})$ corresponding to the respective contact configuration with the contact radius $\tilde{a}$. These basis stresses exhibit the superposition rule
\begin{align}
\tau_B(r;x_i,x_j) + \tau_B(r;x_j,x_k) = \tau_B(r;x_i,x_k) \quad \text{if} \quad x_i > x_j > x_k.
\label{eq_suprule}
\end{align}
Note that for the described basic loading history even the smallest tangential loading will lead to  local slip propagating from the edge of contact, i.e. the case of full stick $c = a$ is only possible for the absence of tangential loading.

In the MDR model the corresponding tangential displacements of the elastic one-dimensional foundation, \cite{PopovHess2014}
\begin{align}
u(x) = \frac{2-\nu}{2G}\int_x^a \frac{r \tau(r)}{\sqrt{r^2-x^2}}~\text{d}r,
\label{eq_ufromtau}
\end{align}
are given by
\begin{align}
u(x) = u_B(x;a,c) \coloneqq \mu \frac{2-\nu}{2-2\nu} \begin{cases}
g(a) - g(c) 	&	\text{if} \quad x \le c \\
g(a) - g(x)		&	\text{if} \quad c < x \le a \\
0,					& \text{else}
\end{cases}
\end{align}
with the equivalent profile
\begin{align}
g(x) \coloneqq x \int_0^x \frac{f'(r)}{\sqrt{x^2-r^2}}~\text{d}r,
\end{align}
where the prime denotes the first derivative. Note that $g$ is the functional relation between indentation depth $\delta$ and contact radius, i.e. \cite{Popovetal2019}
\begin{align}
\delta = g(a),
\end{align}
and that the inverse transform of equation \eqref{eq_ufromtau} reads \cite{Popovetal2019}
\begin{align}
\tau(r) = -\frac{1}{\pi}\frac{4G}{2-\nu}\int_r^a \frac{u'(x)}{\sqrt{x^2-r^2}}~\text{d}x. 
\label{eq_taufromu}
\end{align}
For reasons that will become clear in the following subsections, let us also introduce the function
\begin{align}
F_B(x;a,c) \coloneqq -\frac{\text{d}u_B(x;a,c)}{\text{d}x}\left(\mu \frac{2-\nu}{2-2\nu}\frac{\text{d}g(x)}{\text{d}x}\right)^{-1} = H(x-c) - H(x-a),
\end{align}
with the Heaviside step function $H$. Note that $F_B$ is just a unit rectangle on the interval between $c$ and $a$. 

\subsection{Solution for finite superpositions of basic solutions}\label{sec2_2}

Jäger \cite{Jaeger1993} has shown that the tangential stresses under arbitrary loading histories can always be written as a sum of the basis solutions given above. The sum may be finite or infinite (i.e. an integral). Let us first consider the finite case,
\begin{align}
\tau(r) = \sum_i \kappa_i \tau_B(r;a_i,c_i),
\label{eq_tauforfinite}
\end{align}
with integer coefficients $\kappa_i$ and specific radii $a_i$ and $c_i$; in MDR terms, because of the linearity of the transform \eqref{eq_ufromtau}, we can write
\begin{align}
u(x) = \sum_i \kappa_i u_B(x;a_i,c_i).
\label{eq_uforfinite}
\end{align}
Note that, due to the superposition rule \eqref{eq_suprule}, the decomposition into basis solutions is not unique, which will be used later. 

Now, the construction of that sum depending on the loading history shall not be our concern here (Jäger gave the algorithm and the MDR solution will always be equivalent \cite{PopovHess2014}), but how do we transform between both solutions? 

One direction of the transformation is trivial: Jäger's algorithm will give values of $\kappa_i$, $a_i$ and $c_i$ and equation \eqref{eq_uforfinite} will give the resulting tangential displacements of the elastic foundation in the MDR model. However, how to develop a given function $u(x)$ in a series of basis functions to retrieve the superposition solution? 

As it turns out, that step is also trivial. If we write
\begin{align}
F(x) \coloneqq -\frac{\text{d}u(x)}{\text{d}x}\left(\mu \frac{2-\nu}{2-2\nu}\frac{\text{d}g(x)}{\text{d}x}\right)^{-1},
\label{eq_defF}
\end{align}
because of the linearity of the transform \eqref{eq_ufromtau} and the definition \eqref{eq_defF}, we only have to find the decomposition
\begin{align}
F(x) = \sum_i \kappa_i F_B(x;a_i,c_i), \label{eq_Fxfor finite}
\end{align}
which is a trivial task, because $F_B$ is just a rectangle function.

Note that the link to the MMD can be seen here as well, because the superposition $F(x)$ of rectangular shapes in equation \eqref{eq_Fxfor finite} can be rewritten using the original radial coordinate $r$. The function $F(r)$ that contains a superposition of $r$-dependent rectangular shapes corresponds to the memory diagram introduced in the MMD.

\subsection{Solution for infinite superpositions of basic solutions}\label{sec2_3}

Jäger \cite{Jaeger1993} formulated his contact solution in terms of normal and tangential displacement increments of the rigid indenter. These increments can, but do not necessarily have to, be infinitesimal. Several important special loading scenarios – e.g. pure tangential motion at a fixed indentation depth – lead to non-infinitesimal increments and hence to a discrete sum of basis solutions for the full solution. However, in the general case of arbitrary loading histories the contact solution may be given as an integral of basis solutions. 

To understand the relation between Jäger's solution and the one within the MDR, we first rearrange the finite sum in equation \eqref{eq_tauforfinite}. Let us consecutively number all radii $a_i$ and $c_i$ appearing in the sum into the array $x_k$, with $k$ between zero and some $N$ (depending on the correct superposition solution for the contact problem). Due to the superposition rule \eqref{eq_suprule} we can further split the sum into
\begin{align}
\tau(r) = \sum_i \kappa_i \tau_B(r;a_i,c_i) = \sum_i \kappa_i \sum_{x_k = c_i}^{x_{k+1} = a_i} \tau_B(r;x_{k+1},x_k).
\end{align}
Rearranging this sum, we write
\begin{align}
\tau(r) = \sum_{k = 0}^{N-1} \tilde{\kappa}_k \tau_B(r;x_{k+1},x_k).
\end{align}
In the infinite (i.e. continuous) case the stress increment in the integral is given by
\begin{align}
\text{d}\tau(r) = \tilde{\kappa}(x)\tau_B(r;x + \text{d}x,x).
\end{align}
Moreover, it is for any axisymmetric indenter profile (for brevity, cases which are just zero are omitted)
\begin{align}
\tau_B(r;x + \text{d}x,x) &= \mu \left[p(r;x + \text{d}x) - p(r;x)\right] = \mu \frac{\partial p(r;x)}{\partial x} \text{d}x \\
&= \frac{2\mu G}{\pi (1 - \nu)}\frac{1}{\sqrt{x^2-r^2}}\frac{\text{d}g(x)}{\text{d}x}~\text{d}x,
\end{align}
because the incremental difference between two normal contact configurations with the radii $x$ and $x + \text{d}x$ is equivalent to the infinitesimal indentation by a flat cylindrical punch with the radius $x$ \cite{Mossakovski1963}. Hence, the total shear stress distribution (in the "Jäger picture") is given by
\begin{align}
\tau(r) = \frac{2\mu G}{\pi (1 - \nu)} \int_r^a \frac{\tilde{\kappa}(x)}{\sqrt{x^2-r^2}}\frac{\text{d}g(x)}{\text{d}x}~\text{d}x.
\label{eq_tauforinfinite}
\end{align}
Comparing equations \eqref{eq_taufromu} and \eqref{eq_tauforinfinite} we immediately see that
\begin{align}
\tilde{\kappa}(x) = -\frac{\text{d}u(x)}{\text{d}x}\left(\mu \frac{2-\nu}{2-2\nu}\frac{\text{d}g(x)}{\text{d}x}\right)^{-1} = F(x).
\end{align}
So, the function $F(x)$ defined in equation \eqref{eq_defF} (in the "MDR picture") is just the distribution for the weights $\tilde{\kappa}$ in the superposition solution (i.e. the "Jäger picture")!

\section{Conclusions}\label{sec3}

It has been demonstrated how one can easily switch between two solution procedures for the general axisymmetric tangential contact problem under Hertz-Mindlin assumptions, namely Jäger's superposition of Cattaneo-Mindlin solutions for the elementary loading history and the method of dimensionality reduction. The key to the transition lies in the function $F = \tilde{\kappa}$, defined in equation \eqref{eq_defF}, which gives the superposition weights based on the tangential displacements of the elastic foundation in the MDR model.

The big advantage of the MDR solution is the simplicity of its rules, that makes its implementation almost trivial. For example, all "contact memory" is stored in the tangential displacements $u(x)$ and thus does not have to be tracked separately. However, local contact quantities, like stress distributions, do not exist in the MDR model, but must be retrieved from it via Abel transforms like the one in equation \eqref{eq_taufromu}. With the relations shown above, now there are two possible ways to determine the tangential contact stresses from the MDR solution: direct transformation of the MDR displacements or transition to the Jäger superposition.

If the loading history makes it necessary to use an infinite superposition of Catta\-neo-Mindlin solutions, both ways lead to the same integrals, but in the case of a finite superposition the transition to Jäger's superposition solution does not require any integration at all and will therefore be preferable (especially numerically). The distinction between both cases is trivial based on the function $F$: it is either continuous or stepwise constant.

The given manuscript might leave an impression of "calculating in circles". Naturally, demonstrating the transition between conceptionally very different but equivalent solutions, of course, does not produce new results. However, I am convinced that the mathematics of physics is not supposed to be axiomatic (although some physical principles are obviously more fundamental than others), but rather has a network form; and the more connections between parts of that network are known, the better.

\printbibliography

\end{document}